\documentclass[twocolumn,prc,floatfix,showpacs,preprintnumbers,nofootinbib,%
superscriptaddress]{revtex4}
\usepackage{mathrsfs}
\usepackage{amssymb}
\usepackage{amsmath}
\usepackage{graphicx}

\usepackage{xcolor}
\definecolor{lcolor}{rgb}{0.5,0,0}
\definecolor{citcolor}{rgb}{0,0.3,0.0}

\usepackage[breaklinks,colorlinks,urlcolor=blue,citecolor=citcolor,linkcolor=lcolor]{hyperref}
\usepackage{mciteplus}

\newcommand{\rt}{{\mathbf{r}_T}}

\newcommand{\bt}{{\mathbf{b}_T}}

\newcommand{\Deltat}{{\boldsymbol{\Delta}_T}}

\newcommand{\ud}{\, \mathrm{d}}

\newcommand{\nc}{{N_\mathrm{c}}}

\newcommand{\nr}[1]{(\ref{#1})}

\newcommand{\ra}{R_A}

\newcommand{\gev}{\ \textrm{GeV}}
\newcommand{\fm}{\ \textrm{fm}}

\newcommand{\as}{\alpha_{\mathrm{s}}}

\newcommand{\cN}{\mathcal{N}}

\newcommand{\eq}{Eq.~}

\newcommand{\dsigmap}{{\frac{\ud \sigma^\textrm{p}_\textrm{dip}}{\ud^2 \bt}}}

\newcommand{\dsigma}{{\frac{\ud \sigma_\textrm{dip}}{\ud^2 \bt}}}

\newcommand{\xpom}{{x_\mathbb{P}}}

\newcommand{\Aavg}[1]{\left\langle #1 \right\rangle_\textrm{N}}

\newcommand{\ampli}{{\mathcal{N}}}
\newcommand{\A}{{\mathcal{A}}}

\begin{document}

\author{T. Lappi}
\affiliation{
Department of Physics, %
 P.O. Box 35, 40014 University of Jyv\"askyl\"a, Finland
}

\affiliation{
Helsinki Institute of Physics, P.O. Box 64, 00014 University of Helsinki,
Finland
}
\author{H. M\"antysaari}
\affiliation{
Department of Physics, %
 P.O. Box 35, 40014 University of Jyv\"askyl\"a, Finland
}

\title{
$J/\Psi$ production in ultraperipheral Pb+Pb and p+Pb collisions 
at energies available at the CERN Large Hadron Collider
}


\preprint{}

\begin{abstract}
We compute cross sections for incoherent and coherent diffractive $J/\Psi$ production
in ultraperipheral nucleus-nucleus and proton-nucleus collisions using two 
different dipole models fitted to HERA data. We obtain a reasonably good 
description of the available ALICE data for coherent $J/\Psi$ production 
and present our prediction for the incoherent cross section. We also find 
that while  the normalization of the cross section depends
 quite strongly on the dipole model and vector meson wave function used,
the rapidity dependence is very well constrained.
\end{abstract}

\maketitle

\section{Introduction}

The color glass condensate (CGC) provides a convenient way to describe strongly
interacting systems in the high energy limit, where nonlinear phenomena,
such as gluon recombination, become important. Because the gluon density scales
as $\sim A^{1/3}$, these nonlinearities are 
enhanced when the target is changed from a proton to a heavy nucleus.

The structure of a hadron can be studied accurately in deep inelastic 
scattering (DIS) where a (virtual) photon scatters off the hadron. 
A large amount of precise high energy electron-proton data measured at HERA
has shown that the gluon density inside a proton grows rapidly at small
Bjorken $x$ or,
equivalently, at high energy. These accurate measurements have also been a 
crucial test for the CGC, and recent analyses have confirmed that the CGC
description is consistent with all the available small-$x$ DIS data~ 
\cite{Albacete:2010sy,Rezaeian:2012ji}.

In order to apply  the CGC formalism in the heavy ion environment one would ideally want to 
study DIS off heavy nuclei at high energy. The proposed 
LHeC~\cite{AbelleiraFernandez:2012cc} and
eRHIC~\cite{Boer:2011fh} experiments aim to perform these measurements. 
Before that, one can hope to obtain information about the dense gluonic 
matter in the nucleus by studying, e.g., single~\cite{Tribedy:2011aa,Albacete:2012xq} and 
double inclusive~\cite{Lappi:2012nh,Albacete:2010pg,Stasto:2012ru} 
particle production in proton-nucleus collisions.
These hadronic processes are, however, not ideal for precise studies as
the parton level kinematics is not fixed by the final state particles, in 
contrast to DIS.

Ultraperipheral heavy ion collisions, where two heavy nuclei barely touch 
each other, offer an interesting possibility to study photon-nucleus scattering
without a lepton-ion collider. Recently, the ALICE collaboration has measured
~\cite{Abelev:2012ba}
diffractive vector meson production in these collisions, opening a new 
possibility to study the small-$x$ structure of heavy nuclei. Diffractive
events are especially interesting as they allow one to also study the 
transverse spatial distribution of the gluons in the nuclear wave function,
if the momentum transfer $t$ can be measured.

In this Rapid Communication we present our predictions for coherent and incoherent $J/\Psi$ 
production in heavy ion and pA collisions. In Sec. \ref{sec:dipxs} we first review the 
dipole models used here to compute photon-proton scattering. In Sec. 
\ref{sec:comp} we recall the results for coherent and incoherent vector meson
production in $\gamma$A collisions and discuss how to compute the full 
nucleus-nucleus cross section. Finally in Sec. \ref{sec:res} we present our
results before concluding in Sec. \ref{sec:conclusions}.

\section{Dipole cross sections}
\label{sec:dipxs}

In ultraperipheral nucleus-nucleus collisions
strong interactions are heavily suppressed, and the
electromagnetic interaction is expected to dominate. One can thus
consider one of the nuclei  as a source of (virtual) photons
that scatter off the other nucleus.
In the dipole picture photon-nucleus scattering is described as the 
photon fluctuating into a quark-antiquark color dipole which then scatters
off the target nucleus. The dipole model is valid only when the Bjorken $x$
of the gluon is small; we implement this constraint by computing the cross 
section only when $x<0.02$.

In the literature there are many parametrizations available for the dipole-proton
cross section
\begin{equation}
	\dsigmap(\bt,\rt,\xpom) = 2\cN(\rt,\bt,\xpom),
\end{equation}
where $\cN$ is the imaginary part of the forward dipole-proton scattering 
amplitude, $\rt$ is the transverse
size of the dipole, $\bt$ is the impact parameter of the $\gamma$-p collision
and $\xpom$ is the usual Bjorken variable of DIS in a diffractive event.
The dipole amplitude $\cN$ satisfies the 
BK~\cite{Balitsky:1995ub,*Kovchegov:1999yj,*Kovchegov:1999ua}
evolution equation, and ideally one would want to fit the initial condition
of the BK evolution to the available DIS data (as done in 
Ref.~\cite{Albacete:2010sy}), solve the BK equation and
use the obtained dipole amplitude when computing other observables, such as 
diffractive vector meson production.

However, computing diffractive events requires knowledge about the impact 
parameter dependence of the dipole amplitude. Straightforwardly 
including impact parameter dependence into the BK equation
leads to an unphysical growth of the size of the
proton with the evolution~\cite{GolecBiernat:2003ym} unless this
 is regulated by hand
at the confinement scale~\cite{Berger:2011ew,Berger:2012wx}.
Because of this complication we use in this work 
two phenomenological dipole cross section parametrizations
that include a realistic impact parameter dependence.
One is the 
IIM~\cite{Iancu:2003ge} dipole cross section which is a 
parametrization including the most important features 
of BK evolution. The detailed expression for the
dipole cross section can be found in Ref.~\cite{Iancu:2003ge};
we use here the values of the parameters from the newer
fit to HERA data including charm from Ref.~\cite{Soyez:2007kg}.
The second parametrization used here is a factorized approximation of 
the IPsat model with an eikonalized DGLAP-evolved gluon distribution 
\cite{Kowalski:2003hm,Kowalski:2006hc}.

In the IIM model the impact parameter dependence is explicitly factorized as
\begin{equation}\label{eq:factbt}
\dsigmap(\bt,\rt,x) = 2T_p(\bt) \ampli(\rt,x),
\end{equation}
We take, following Ref.~\cite{Marquet:2007nf}, a Gaussian profile for the
proton impact parameter profile function:
$T_p(\bt) = \exp\left(-b^2/2 B_p\right)$ with 
$B_p=5.59\gev^{-2}$. 

In the IPsat model the impact parameter dependence is
included in the saturation scale as
\begin{equation}\label{eq:unfactbt}
\dsigmap(\bt,\rt,x)
 = 2\,\left[ 1 - \exp\left(- r^2  F(x,r) T_p(\bt)\right) 
\right],
\end{equation}
denoting $r=|\rt|$. Here $T_p(\bt)$ is the same impact parameter profile function as above, but
the fitted value for the proton shape is  $B_p=4.0\gev^2$ (see 
Ref.~\cite{Lappi:2010dd} for a discussion about the different numerical value)
and $F$ is proportional to the 
DGLAP evolved gluon distribution~\cite{Bartels:2002cj},
\begin{equation}
F(x,r) = 
\frac{1}{2 \pi B_p}
\frac{ \pi^2 }{2 \nc} \as \left(\mu_0^2 + \frac{C}{r^2} \right) 
x g\left(x,\mu_0^2 + \frac{C}{r^2} \right),  
\label{eq:BEKW_F}
\end{equation}
with $C$ chosen as 4 and $\mu_0^2=1.17\gev^2$ resulting from the 
fit~\cite{Kowalski:2006hc}. Following Ref.~\cite{Lappi:2010dd} we
replace \eq\nr{eq:unfactbt} by the factorized approximation
\begin{equation}\label{eq:BEKWfact}
\dsigmap(\bt,\rt,x)
 \approx  2 T_p(\bt) \,\left[ 1 - \exp\left(- r^2  F(x,r)\right)
\right],
\end{equation}
using the same $F(x,r)$ defined in \eq\nr{eq:BEKW_F}. This approximation,
denoted here as ``fIPsat'',
brings the IPsat parametrization to the form \eq\nr{eq:factbt}
with  $\ampli(r,x)=\left[ 1 - \exp\left(- r^2  F(x,r)\right)\right]$.
It was shown in Ref.~\cite{Lappi:2010dd} that the fIPsat 
parametrization also describes the HERA $J/\Psi$ data accurately.

\section{Diffractive cross section in ultraperipheral collisions}
\label{sec:comp}

In this work we consider both coherent and incoherent diffractive vector meson 
production. In a coherent process the nucleus off which the photon scatters 
remains intact, whereas in incoherent diffraction the nucleus is allowed to 
break up. The event is still diffractive (there is a rapidity gap) as long as 
there is no exchange of color charge.

The cross section for quasielastic (coherent+incoherent) vector meson 
production in nuclear DIS is (see, e.g.,~\cite{Kowalski:2006hc}) 
\begin{equation} \label{eq:xsec}
\frac{\ud \sigma^{\gamma^* A \to V A }}{\ud t} 
= \frac{R_g^2(1+\beta^2)}{16\pi} \Aavg{|\A(\xpom,Q^2,\Deltat)|^2},
\end{equation}
where $-Q^2$ is the virtuality of the photon.
The coherent cross section is obtained by averaging the amplitude
before squaring it, $|\Aavg{\A}|^2$, and the incoherent one 
is given by the variance $\Aavg{|\A|^2}-|\Aavg{\A}|^2$,
(see Refs.~\cite{Lappi:2010dd,Toll:2012mb})
where
\begin{equation} \label{eq:aavg}
\Aavg{\mathcal{O}(\{ \bt_i \})} 
\equiv \int \prod_{i=1}^{A}\left[ \ud^2 \bt_i T_A(\bt_i) \right] 
\mathcal{O}(\{ \bt_i \})
\end{equation}
is the average over the positions of the nucleons in the nucleus.
Here $T_A$ is the Woods-Saxon distribution with nuclear radius 
$\ra = (1.12 A^{1/3}-0.86 A^{-1/3})\fm $ and surface thickness 
$d=0.54\fm$.

The factor $1+\beta^2$ accounts for the
real part of the scattering amplitude and the factor $R_g^2$ corrects
for the skewedness effect, i.e. that the gluons in the target are probed at 
slightly different $\xpom$~\cite{Shuvaev:1999ce,*Martin:1999wb}. 
For these corrections
we follow the prescription of Ref.~\cite{Watt:2007nr}, taking them 
as
\begin{eqnarray}
\beta &=& \tan \frac{\pi \lambda}{2}
\\  
R_g &=& \frac{2^{2 \lambda+3}}{\sqrt{\pi}}\frac{\Gamma(\lambda + 5/2)}{\Gamma(\lambda+4)} 
\quad \textrm{ with}
\\
\lambda &=& \frac{\partial \ln \A}{\partial \ln 1/\xpom}.
\end{eqnarray}
We calculate, as in Ref.~\cite{Lappi:2010dd},
the correction terms from the energy dependence of the nucleon
scattering amplitudes and use the same values for the nucleus at the 
same $Q^2,\xpom$. 
The real part and skewedness corrections, especially $R_g$, are 
a significant  factor in the absolute normalization of the cross section and are 
necessary for an agreement with HERA data.

The imaginary part of the scattering amplitude, $\A$, is the Fourier-transform 
of the dipole-target cross section $\sigma_\text{dip}$ from impact parameter
$\bt$ to momentum transfer $\Deltat$, contracted with the overlap between the 
vector meson and virtual photon wave functions:
\begin{multline}\label{eq:ampli}
\A(\xpom,Q^2,\Deltat) 
= \int \ud^2 \rt \int \frac{\ud z}{4\pi} \int \ud^2 \bt 
\\
\times [\Psi_V^* \Psi](r,Q^2,z)
e^{-i \bt \cdot  \Deltat}  
\dsigma(\bt,\rt,\xpom),
\end{multline}
where we have followed the normalization convention of Ref.~\cite{Kowalski:2006hc}.
For the virtual photon--vector meson wavefunction overlap we
use the ``boosted Gaussian'' and ``gaus-LC'' parametrizations
 from Ref.~\cite{Kowalski:2006hc}.

Assuming a large and smooth nucleus the averaged amplitude required to
compute coherent $J/\Psi$ production reads~\cite{Kowalski:2003hm}
\begin{multline}\label{eq:cohampli}
\Aavg{\A(\xpom,Q^2,\Deltat) }
= \int \frac{\ud z}{4\pi} \ud^2 \rt  \ud^2 \bt e^{-i \bt \cdot  \Deltat}  
\\
 \times [\Psi_V^*\Psi](r,Q^2,z)
\,  2 \left[ 1-\exp\left\{ - 2 \pi B_p A T_A(b) \ampli(r,\xpom) \right\} \right].
\end{multline}
At large $-t=\Deltat^2$  the cross
section is almost purely incoherent. Thus the incoherent cross section
can at large $|t|$ be computed as the total quasielastic cross section,
 by first squaring and then averaging the amplitude.
The result is derived, e.g., in Ref.~\cite{Lappi:2010dd} and reads
\begin{multline}\label{eq:amplisqn1}
\left \langle \left| \mathcal{A}_{q\bar{q}}\right|^2(\xpom, Q^2, \Deltat) \right\rangle_N
=
16 \pi B_p A \int \ud^2 \bt 
\\ \times \int \ud^2 \rt \ud^2 \rt' \frac{\ud z}{4\pi} \frac{\ud z'}{4\pi} [\Psi^*_V\Psi](r,Q^2,z) [\Psi_V^*\Psi](r',Q^2,z') \\
\\ \times 
e^{-B_p \Deltat^2}
e^{-2 \pi B_p A T_A(b)
\left[ \ampli(r) + \ampli(r') \right] } 
\\ \times
\left( \frac{\pi B_p \ampli(r)\ampli(r') T_A(b) }
  {1 - 2 \pi B_p T_A(b)\left[ \ampli(r) + \ampli(r') \right] } \right).
\end{multline}

Following Ref.~\cite{Bertulani:2005ru} we factorize the diffractive vector 
meson production cross section in nucleus-nucleus (or proton-nucleus) 
collisions to the product of the equivalent photon flux 
generated by one of the nuclei and the photon-nucleus cross section:
\begin{equation}
	\sigma^{AA\to J/\Psi A} = \int \ud \omega \frac{n(\omega)}{\omega} 
		\sigma^{\gamma A \to J/\Psi A}(\omega).
\end{equation}
Here $\sigma^{\gamma A \to J/\Psi A}$ is the diffractive photon-nucleus cross section,
$\omega=(M_V/2)e^y$ is the energy of the photon in the collider frame and 
$M_V$ and $y$ are the vector meson mass and rapidity. The explicit
expression for the  photon flux $n(\omega)$ (integrated over the
impact parameter of the AA-collision ${\bf b}_T^{AA} > 2R_A$) can be found 
in Ref.~\cite{Bertulani:2005ru}. 
In nucleus-nucleus collisions both nuclei can act as a source of photons
that scatter off the other nucleus:
\begin{equation}
	\frac{\ud \sigma^{A_1A_2\to J/\Psi A}}{\ud y} = n^{A_2}(y) \sigma^{\gamma A_1}(y) + n^{A_1}(-y) \sigma^{\gamma A_2}(-y).
\end{equation}

In proton-nucleus collisions the photon flux generated by a nucleus is computed
requiring that the impact parameter is larger than $R_A$.
The proton can also act as a photon source, and the photon flux generated
by a proton is computed as in Ref.~\cite{Bertulani:2005ru}. As the photon flux
is proportional to the charge squared, the process where the photon is emitted
from the nucleus dominates.

The kinematics of diffractive vector meson production is such 
that the gluon $\xpom$ probed by the real photon is
$\xpom=M_V e^{-y} /\sqrt{s_\text{NN}} $. At forward and backward 
rapidities we have two different contributions: either a small-$x$ photon 
scatters off a large-$x$ gluon or vice versa. At midrapidity we only probe
small-$x$ structure of the nucleus. Our results should be most reliable in that
region. At the LHC $\sqrt{s_\text{NN}}=2.76$ TeV,
and for $J/\Psi$ production $\xpom \approx 0.001$ at $y=0$.

\section{Results and discussion}
\label{sec:res}

\begin{figure}[tb]
\begin{center}
\includegraphics[width=0.49\textwidth]{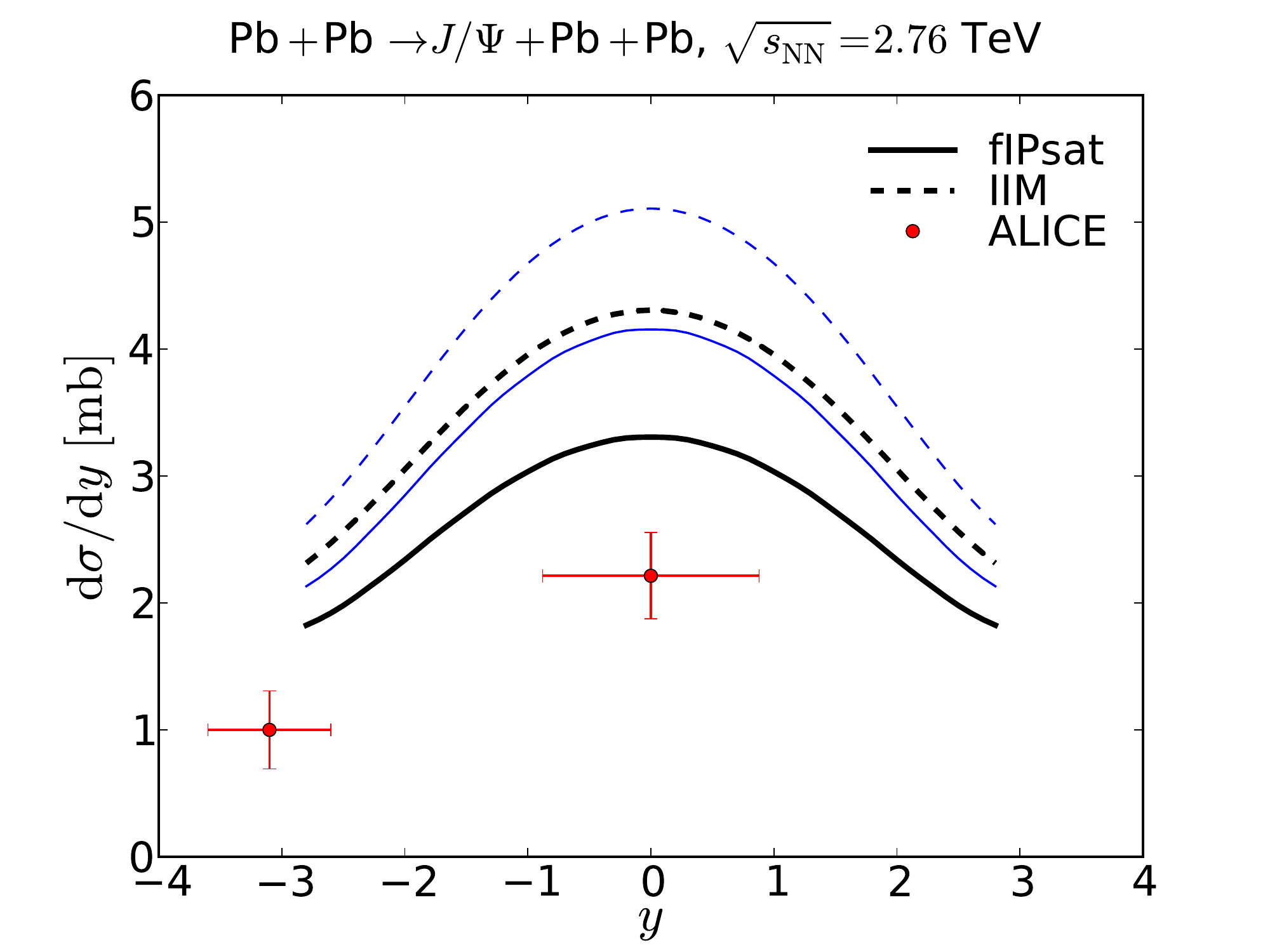}
\end{center}
\caption{
The coherent diffractive $J/\Psi$ photoproduction ($Q^2=0$ GeV$^2$) cross section in lead-lead collisions at $\sqrt{s_\mathrm{NN}}=2.76$ TeV computed using fIPsat and IIM parametrizations and Boosted Gaussian (thin blue lines) and Gaus-LC (thick black lines) wavefunctions compared with the ALICE data~\cite{Abelev:2012ba,MayerCracowTalk:2013}.
}\label{fig:coherent}
\end{figure}

\begin{figure}[tb]
\begin{center}
\includegraphics[width=0.49\textwidth]{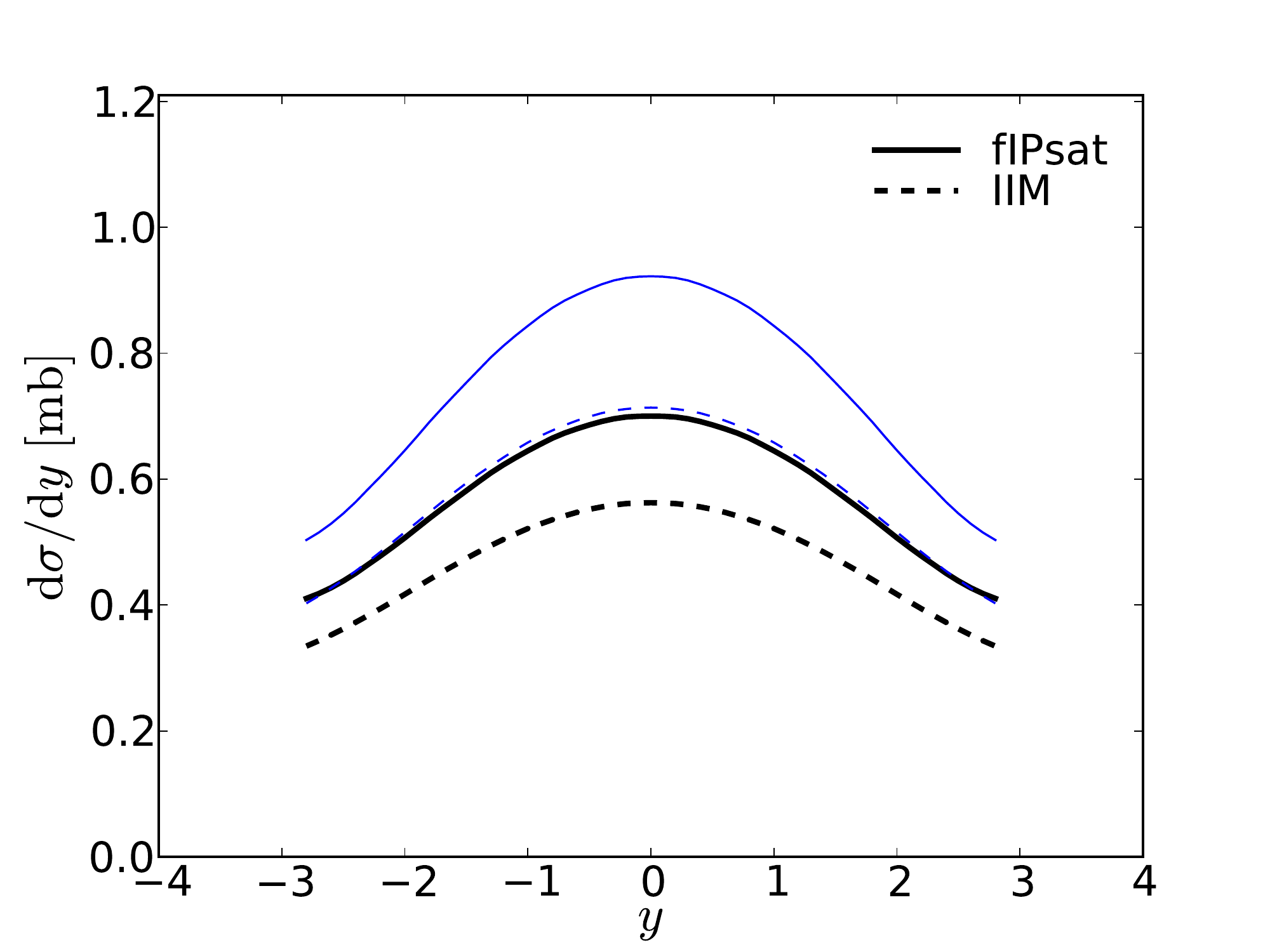}
\end{center}
\caption{
The incoherent diffractive $J/\Psi$ photoproduction cross section in lead-lead collisions at $\sqrt{s_\mathrm{NN}}=2.76$ TeV computed using fIPsat and IIM parametrizations and Boosted Gaussian (thin blue lines) and Gaus-LC (thick black lines) wavefunctions.
}\label{fig:incoherent}
\end{figure}

\begin{figure}[tb]
\begin{center}
\includegraphics[width=0.49\textwidth]{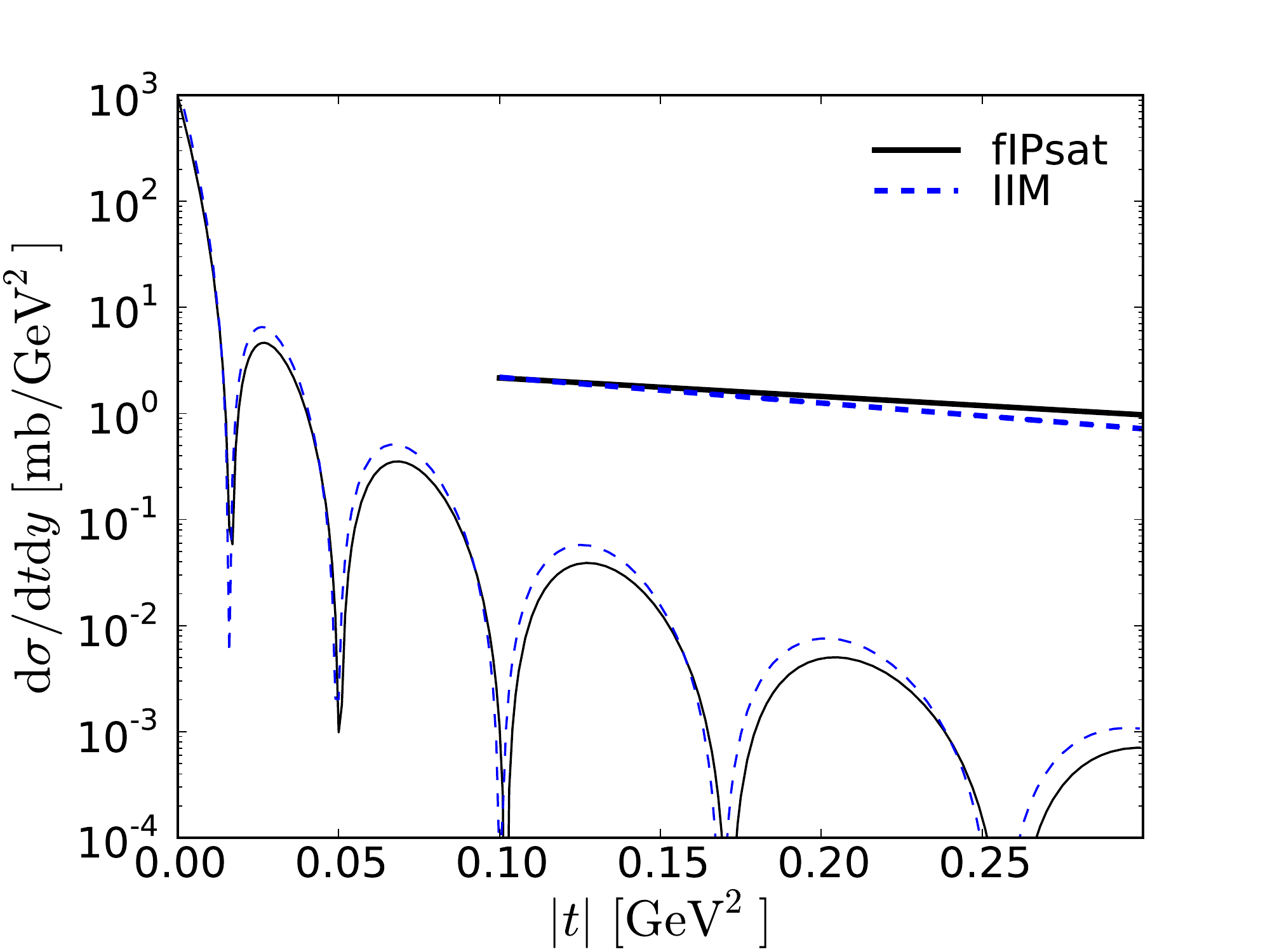}
\end{center}
\caption{
The coherent (thick lines) and incoherent (thin lines) diffractive $J/\Psi$ photoproduction cross section in lead-lead collision at $\sqrt{s_{\text{NN}}}=2.76$ TeV as a function of momentum transfer $t$ at midrapidity $y=0$ using the Gaus-LC wave function. 
}\label{fig:incoh-t}
\end{figure}

\begin{figure}[tb]
\begin{center}
\includegraphics[width=0.49\textwidth]{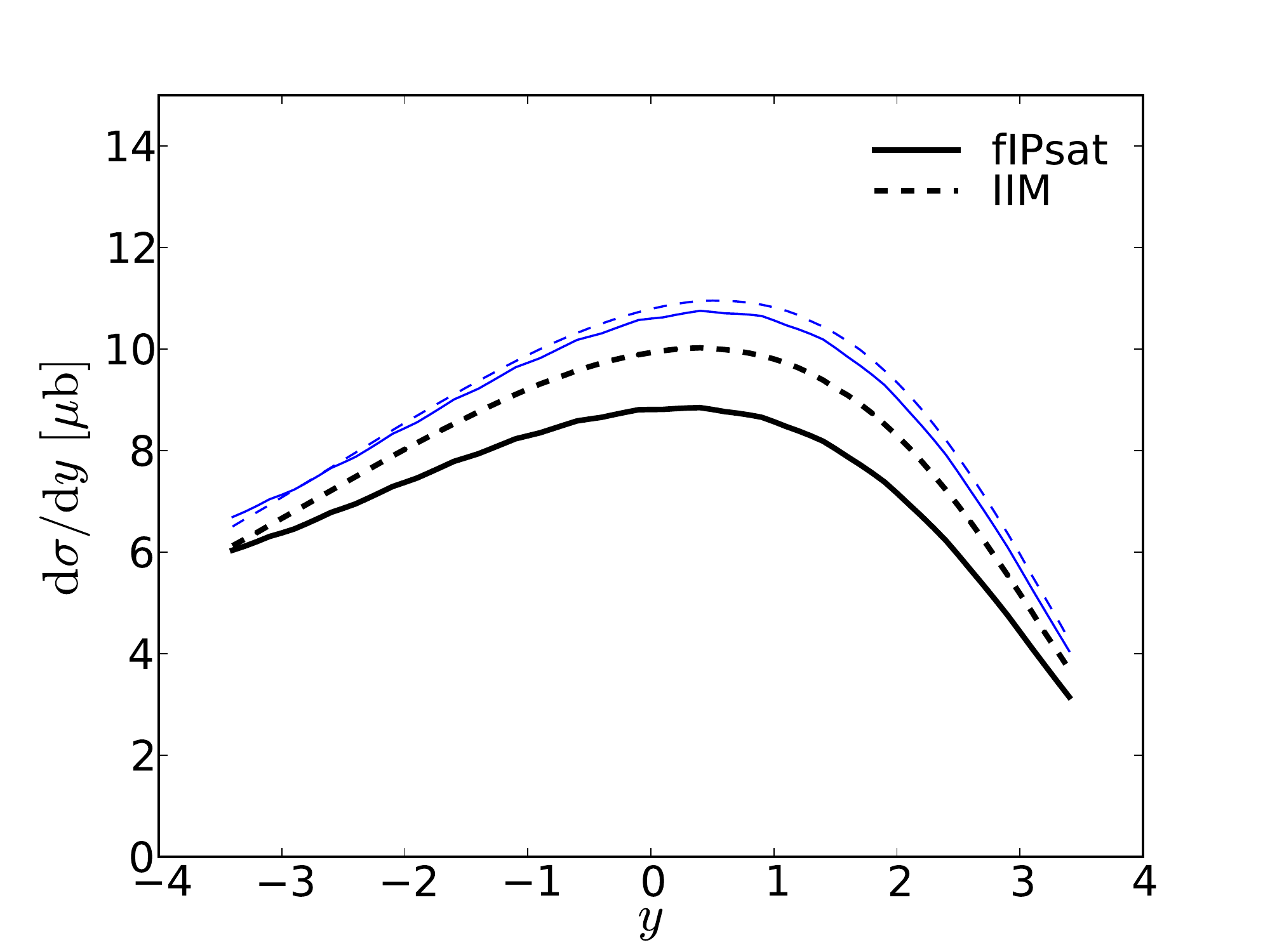}
\end{center}
\caption{
The diffractive $J/\Psi$ photoproduction cross section in proton-lead collisions at $\sqrt{s_\text{NN}}=5.02$ TeV computed using fIPsat and IIM parametrizations and boosted Gaussian (thin blue lines) and Gaus-LC (thick black lines) wave functions. The proton is moving in the negative $y$ direction.
}\label{fig:pa}
\end{figure}

\begin{figure}[tb]
\begin{center}
\includegraphics[width=0.49\textwidth]{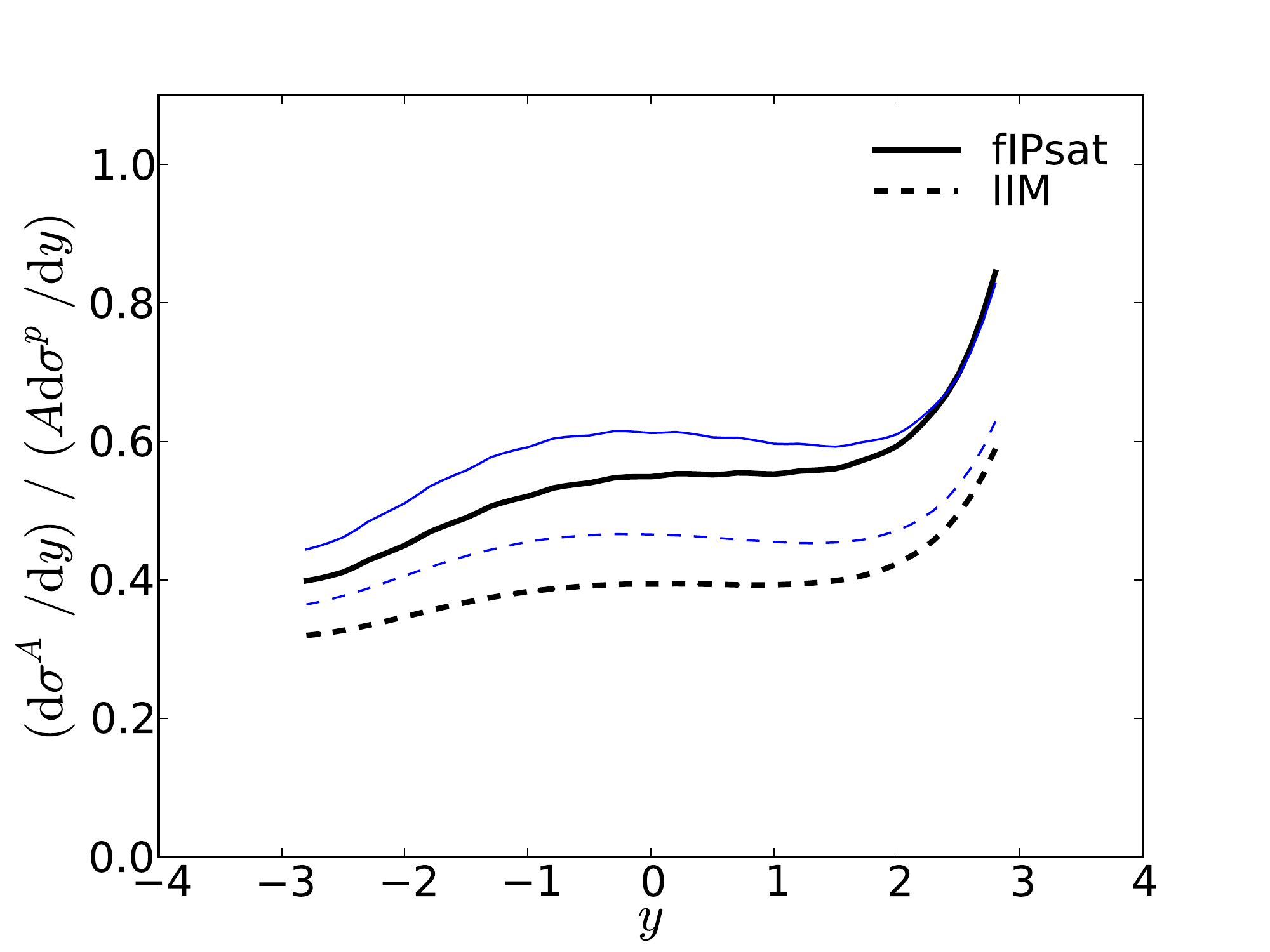}
\end{center}
\caption{
Incoherent diffractive $J\Psi$ photoproduction cross section in lead-lead collision normalized by coherent $J/\Psi$ production in pA (nuclear transparency ratio) at $\sqrt{s_\text{NN}}=2.76$ TeV computed using fIPsat and IIM parametrizations and boosted Gaussian (thin blue lines) and Gaus-LC (thick black lines) wave functions. The proton is moving in the negative $y$ direction.
}\label{fig:rpa}
\end{figure}

The ALICE Collaboration has measured the coherent $J/\Psi$ photoproduction 
 cross section $\ud \sigma/\ud y$ at a relatively large rapidity $|y|\sim 3$~\cite{Abelev:2012ba}. Recently preliminary results at midrapidity $y=0$ were also published~\cite{MayerCracowTalk:2013}. The comparison between our results and the ALICE data is shown in Fig.~\ref{fig:coherent}. The PHENIX Collaboration has also measured the ultraperipheral $J/\Psi$ cross section
in gold-gold collisions at $\sqrt{s_\text{NN}}=200$ GeV and $y=0$. For these kinematics the
fIPsat dipole cross section with the gaus-LC wavefunction 
gives the result $109 \mu \textup{b}$, compared to $76 \pm 34 \mu \textup{b}$
measured by PHENIX~\cite{Afanasiev:2009hy}. 
The ALICE data seems to favor the fIPsat over the IIM model. This is perhaps not surprising. The most important difference between the two dipole models for this purpose is the different 
impact parameter dependence. 
The IIM value $B_p=5.59\gev^{-2}$ comes from a fit to inclusive data and is close to the measured 
value for inclusive diffraction. The HERA data for diffractive $J/\Psi$ production~\cite{Chekanov:2002xi,Aktas:2005xu}  has, however, a smaller $B_p$, which is reflected in the IPsat parametrization. Thus we consider results obtained using the fIPsat model more reliable.

Our results slightly overshoot the data, but the rapidity dependence comes out correctly. For example the fIPsat model with Gaus-LC wavefunction is above all data points by a factor $\sim 1.4$.
We consider the agreement relatively
good given the simplicity of the parametrizations and the fact that no nuclear data
was used to constrain the models. 
 We emphasize that the parametrizations used are exactly the
same as in Ref.~\cite{Lappi:2010dd}, which predates the ALICE data. 
Recall that both dipole models and wavefunctions used here give good descriptions of the 
HERA data. We suspect that the main reason for the larger normalization lies in the 
skewedness correction. The correction is larger here than for HERA kinematics because of the
larger $x$ probed, making it less reliable.

The difference between the two wavefunctions is largest at $Q^2=0$ which is the case here. The ALICE data seems to favor the Gaus-LC wavefunction, and thus we consider the fIPsat dipole model and the Gaus-LC wavefunction to be the most reliable combination. At $|y|\gtrsim 2$, $\sqrt{s_\text{NN}}=2.76$ TeV, we are probing gluons with $\xpom \gtrsim 0.01$, and in that region our parametrizations for the dipole amplitude are not valid any more. In addition, the real part and skewedness corrections in total become of the order $2$, making them less reliable.
 Nevertheless, all wavefunctions and dipole models consistently give $\ud \sigma/\ud y|_{y=0}\,/\,\ud \sigma/\ud y|_{y=2} = 1.41 \textrm{--} 1.46$. Thus the prediction for the rapidity dependence is much more robust than for the absolute normalization.

We then present our predictions for the incoherent diffractive vector meson cross section in Fig.~\ref{fig:incoherent}. Again the different models give a quite different overall
normalization, but a very similar rapidity dependence.
Notice that now the absolute normalization is larger in the fIPsat model.
This is due to the different different impact parameter profiles,
we refer the reader to Ref.~\cite{Lappi:2010dd} for a more detailed discussion.
Again most of the difference cancels in the ratio $\ud \sigma/\ud y|_{y=0}\,/\,\ud \sigma/\ud y|_{y=2}$ which is now $1.35 \textrm{--} 1.43$, so the energy dependence is very similar in coherent and incoherent scattering. 

Our result for the incoherent vector meson production from Ref.~\cite{Lappi:2010dd} is not valid at small $|t|$. However, we expect to get a realistic estimate for the total incoherent cross section by integrating the differential cross section starting from the value of $|t|$  where incoherent and coherent cross sections are equal.
The error made is small, 
 parametrically a factor $\sim e^{-|t_\textup{min}|B_p }$ with 
$|t_\textup{min}|\sim 1/\ra^2$ and numerically $\lesssim 10\%$.

In Fig. \ref{fig:incoh-t} we present predictions for the $t$ distribution of diffractive $J/\Psi$ photoproduction at midrapidity where $\xpom\approx 0.001$. Note that the $t$ slope of the incoherent cross section directly measures the spatial distribution of gluons inside a \emph{nucleon}, because the $t$-dependence of the incoherent cross section is $\sim\exp(B_p t)$~\cite{Lappi:2010dd}.

Our pA results are shown in Figs.~\ref{fig:pa} and \ref{fig:rpa}. In Fig. \ref{fig:pa} we show the rapidity dependence of the diffractive $J/\Psi$ cross section (the photon-nucleus scattering is required to be coherent).
Now the difference between the models is reduced as the dominant process is  photon-proton scattering where the models are constrained by HERA data.
Finally, in Fig.~\ref{fig:rpa} we compute the incoherent $J/\Psi$ photoproduction cross section in $AA$ collisions divided by $A$ times the diffractive $J/\Psi$ production cross section in pA collisions at the same $\sqrt{s_\mathrm{NN}}$. Since ultraperipheral proton-nucleus collisions are mostly photon-proton collisions, this is a 
``nuclear transparency'' ratio that measures the absorption of the dipole as it propagates through the nucleus, see, e.g.,~\cite{Kopeliovich:2000ra,Lappi:2010dd}.

\section{Conclusions}
\label{sec:conclusions}
We have computed coherent and incoherent diffractive $J/\Psi$ photoproduction cross sections in ultraperipheral heavy ion collisions. The only inputs to our calculation come from fits to HERA data and standard nuclear geometry.
Especially the rapidity dependence agrees relatively well with the published ALICE result, considering the rather large dependence on the details of the dipole model and the vector meson light cone wavefunction.
The normalization of the data favors the  fIPsat parametrization, where the proton diffractive slope is constrained by the HERA diffractive $J/\Psi$ data.
We also find that the different parametrizations, each known to fit HERA data well, yield significantly different normalizations for the cross section leaving the overall rapidity dependence  very similar, 
with  $\ud \sigma/\ud y|_{y=0}\,/\,\ud \sigma/\ud y|_{y=2} = 1.41 \textrm{--} 1.46$ for the coherent and 
$\ud \sigma/\ud y|_{y=0}\,/\,\ud \sigma/\ud y|_{y=2} = 1.35 \textrm{--} 1.43$ for the incoherent scattering.
A similar conclusion was found in Ref.~\cite{Lappi:2010dd} for photon-nucleus scattering.
We also present predictions for diffractive $J/\Psi$ production in pA collisions.

\section*{Acknowledgements}
We thank J. Nystrand for suggesting this topic for us.
This work has been supported by the Academy of Finland, Project No.
133005. H.M. is supported by the Graduate School of Particle and Nuclear 
Physics.

\bibliography{../../../refs}
\bibliographystyle{JHEP-2modM}

\end{document}